\documentclass[12pt]{article}
\usepackage{verbatim,a4wide,epsfig,amsmath,chicago,url}

\newcommand{\E}{\mbox{E}}
\newcommand{\Var}{\mbox{Var}}
\newcommand{\SD}{\mbox{SD}}

\newcommand{\lp}{\left(}
\newcommand{\rp}{\right)}
\newcommand{\hatb}{\hat b_{UCV}}

\def\conp{\buildrel p\over \longrightarrow}
\def\cond{\buildrel {\cal D}\over \longrightarrow}
\def\ra{\rightarrow}

\def\medi{\medskip\noindent}
\def\beqn{\begin{eqnarray*}}
\def\eeqn{\end{eqnarray*}}
\def\beq{\begin{eqnarray}}
\def\eeq{\end{eqnarray}}
\def\cv{cross-validation~}
\def\cvp{cross-validation}

\begin{document}
\title{\bf Indirect Cross-validation for Density Estimation}
\author{Olga Y. Savchuk, Jeffrey D. Hart, Simon J. Sheather}
%\date{\today}
\date{}
\maketitle

\begin{abstract}
A new method of bandwidth selection for kernel density estimators
is proposed. The method, termed {\it indirect cross-validation}, or
ICV, makes use of so-called {\it selection} kernels. Least squares
cross-validation (LSCV) is used to select the bandwidth of a
selection-kernel estimator, and this bandwidth is appropriately rescaled
for use in a Gaussian kernel estimator. The proposed
selection kernels are linear
combinations of two Gaussian kernels, and need not
be unimodal or positive. Theory is developed showing that the relative
error of ICV bandwidths can converge to 0 at a rate of $n^{-1/4}$,
which is substantially better than the $n^{-1/10}$ rate of
LSCV. Interestingly, the selection kernels that are best for purposes
of bandwidth selection are very poor if used to actually estimate the
density function. This property appears to be part of the larger and
well-documented paradox to the effect that ``the harder the estimation
problem, the better \cv performs.'' The ICV method uniformly outperforms
LSCV in a simulation study, a real data example, and a simulated
example in which bandwidths are chosen locally.

\end{abstract}

\section{Introduction}

Let $X_1,\ldots,X_n$ be a random sample from an unknown density
$f$. A kernel density estimator of $f(x)$ is
\begin{equation}
\label{eq:KDE} \hat f_h(x)=\frac{1}{nh}\sum_{i=1}^n
K\Bigl(\frac{x-X_i}{h}\Bigr),
\end{equation}
where $h>0$ is a smoothing parameter, also known as the bandwidth,
and $K$ is the kernel, which is generally chosen to be a unimodal
probability density function that is
symmetric about zero and has finite variance.  A popular choice
for $K$ is the Gaussian kernel: $\phi(u)=(2\pi)^{-1/2}\exp(-u^2/2)$.
To distinguish between estimators with different kernels, we shall
refer to estimator \eqref{eq:KDE} with given kernel $K$ as a {\it
  $K$-kernel estimator}. Choosing an appropriate bandwidth is vital
for the good performance of a kernel estimate.  This paper is
concerned with a new method of data-driven bandwidth selection that we
call {\it indirect cross-validation} (ICV).

Many data-driven methods of bandwidth selection have been proposed. The
two most widely used are least squares cross-validation,
proposed independently by~\citeN{Rudemo:LSCV}
and~\citeN{Bowman:LSCV}, and the~\citeN{Sheather:PI} plug-in
method. Plug-in produces more stable bandwidths than does
cross-validation, and hence is the currently more popular
method. Nonetheless, an argument can be made for cross-validation
since it requires fewer assumptions than plug-in and works well when
the density is difficult to estimate; see \citeN{Loader:Classical}.
%The main idea of the plug-in method is to estimate
%$R(f^{\prime\prime})$ and plug it into expression~\eqref{eq:h_n}.
A survey of bandwidth selection methods is given by~\citeN{Jones:survey}.

A number of modifications of LSCV has been proposed in an attempt to
improve its performance. These include the biased \cv method
of~\citeN{Scott:UCV}, a method of~\citeN{Chiu:CV}, the
trimmed \cv of~\citeN{Feluch:spacing}, the modified \cv
of~\citeN{Stute}, and the method of~\citeN{Ahmad} based on kernel
contrasts.
%A survey of bandwidth selection methods up to 1996 is provided
%by~\citeN{Jones:survey}.
The ICV method is similar in spirit to one-sided cross-validation
(OSCV), which is another modification of cross-validation proposed in
the regression context by~\citeN{HartYi}. As in OSCV,
ICV initially chooses the bandwidth of an $L$-kernel
estimator using least squares cross-validation. Multiplying the
bandwidth chosen at this initial stage by a known constant results in
a bandwidth, call it $\hat h_{ICV}$, that is appropriate for use in a
Gaussian kernel estimator.

A popular means of judging a kernel
estimator is the mean integrated squared error, i.e.,
$MISE(h)=E\left[ISE(h)\right]$, where
$$
ISE(h)=\int_{-\infty}^\infty \left(\hat f_h(x)-f(x)\right)^2\,dx.
$$
Letting $h_0$ be the bandwidth that minimizes $MISE(h)$ when the
kernel is Gaussian, we will show that the
mean squared error of $\hat h_{ICV}$ as an estimator of $h_0$
converges to 0
at a faster rate than that of the ordinary
LSCV bandwidth. We also describe an unexpected bonus associated with
ICV, namely that, unlike LSCV, it is robust to rounded data. A fairly
extensive simulation study and two data analyses confirm that ICV
performs better than ordinary cross-validation in finite samples.

\section{Description of indirect cross-validation}
%e propose modifying the LSCV criterion function by using a
%ifferent kernel at the cross-validation stage than at the
%stimation stage. Initially we used kernels (at the CV stage) that
%ownweighted the smallest spacings in the data, but later learned
%hat some other kernels actually worked better.
We begin with some notation and definitions that will be used
subsequently. For an arbitrary function $g$, define
\[
R(g)=\int g(u)^2\,du,\quad \mu_{jg}=\int u^j g(u)\,du.
\]
The LSCV criterion is given by
$$
LSCV(h)=R(\hat f_h)-\frac{2}{n}\sum_{i=1}^n\hat f_{h,-i}(X_i),
$$
where, for $i=1,\ldots,n$, $\hat f_{h,-i}$ denotes a kernel estimator
using all the original observations except for
$X_i$. When $\hat f_h$ uses kernel $K$, $LSCV$ can be
written as
\beq\label{eq:LSCV1}
LSCV(h)&=&\frac{1}{nh}R(K)+\frac{1}{n^2h}\sum_{i\neq
j}\int K(t)K\Bigl(t+\frac{X_i-X_j}{h}\Bigr)\,dt\notag\\
&&-\frac{2}{n(n-1)h}\sum_{i\neq j}K\Bigl(\frac{X_i-X_j}{h}\Bigr).
\eeq
It is well known that $LSCV(h)$ is an unbiased estimator of
$MISE(h)-\int f^2(x)\,dx$, and hence the minimizer of $LSCV(h)$ with
respect to $h$ is denoted $\hat h_{UCV}$.

\subsection{The basic method}\label{sec:meth}
Our aim is to choose the bandwidth of a {\it second order} kernel
estimator. A second order kernel
integrates to 1, has first moment 0, and
finite, nonzero second moment. In principle our method can be used to choose
the bandwidth of any second order kernel estimator, but in this
article we restrict attention to $K\equiv \phi$, the Gaussian
kernel. It is well known that a $\phi$-kernel estimator has asymptotic
mean integrated squared error (MISE) within 5\% of the minimum among
all positive, second order kernel estimators.

Indirect cross-validation may be described as follows:
\begin{itemize}
\item Select the bandwidth of an $L$-kernel estimator using least
  squares cross-val- idation, and call this bandwidth $\hat
  b_{UCV}$. The kernel $L$ is a second order kernel that is a linear
  combination of two Gaussian kernels, and will be discussed in detail
  in Section \ref{sec:wildk}.
\item Assuming that the underlying density $f$ has second derivative which
is continuous and square integrable, the
  bandwidths $h_n$ and $b_n$ that asymptotically minimize the $MISE$ of
  $\phi$- and $L$-kernel estimators, respectively,  are related as
  follows:
\beq\label{eq:hnbn}
h_n=\left(\frac{R(\phi)\mu_{2L}^2}{R(L)\mu_{2\phi}^2}\right)^{1/5}b_n\equiv
Cb_n.
\eeq
\item Define the indirect cross-validation bandwidth by $\hat
  h_{ICV}=C\hat b_{UCV}$. Importantly, the constant $C$ depends on no
  unknown parameters. Expression (\ref{eq:hnbn}) and existing
  cross-validation theory suggest that $\hat h_{ICV}/h_0$ will at least
  converge to 1 in probability, where $h_0$ is the minimizer of
  $MISE$ for the $\phi$-kernel estimator.
\end{itemize}

Henceforth, we let $\hat h_{UCV}$ denote the bandwidth that minimizes
$LSCV(h)$ with $K\equiv \phi$. Theory of~\citeN{H&M:Extent}
and~\citeN{Scott:UCV}
shows that the relative error $(\hat h_{UCV}-h_0)/h_0$
converges to 0 at the rather disappointing rate of $n^{-1/10}$. In
contrast, we will show that $(\hat h_{ICV}-h_0)/h_0$ can converge to 0
at the rate $n^{-1/4}$. Kernels $L$ that are sufficient for this
result are discussed next.

\subsection{Selection kernels}\label{sec:wildk}
We consider the family of kernels ${\cal
  L}=\{L(\,\cdot\,;\alpha,\sigma): \alpha\ge0,\sigma>0\}$, where, for
all $u$,
\begin{equation}
\label{eq:L}
L(u;\alpha,\sigma)=(1+\alpha)\phi(u)-\frac{\alpha}{\sigma}\phi\left(\frac{u}{\sigma}\right).
\end{equation}
Note that the Gaussian kernel is a special case of~\eqref{eq:L} when
$\alpha=0$ or $\sigma=1$. Each member of ${\cal L}$ is symmetric about
0 and such that $\mu_{2L}=\int
u^2L(u)\,du=1+\alpha-\alpha\sigma^2$. It follows that kernels in
${\cal L}$ are second order, with the exception of those for which
$\sigma=\sqrt{(1+\alpha)/\alpha}$.

The family ${\cal L}$ can be partitioned into three families:
${\cal L}_1$,  ${\cal L}_2$ and  ${\cal L}_3$. The first of these is
$\mathcal L_1=\bigl\{L(\cdot;\alpha,\sigma):
\alpha>0, \sigma<\frac{\alpha}{1+\alpha}\bigr\}$. Each kernel in ${\cal
  L}_1$ has a negative dip centered at $x=0$.
%and hence will
%downweight the smallest spacings in a CV curve (see expression
%(\ref{eq:LSCV1})).
For $\alpha$ fixed,
the smaller $\sigma$ is, the more extreme the
dip; and for fixed $\sigma$, the larger $\alpha$ is, the more
extreme the dip. The kernels in $\mathcal L_1$ are ones that
``cut-out-the-middle.''
%some examples of which are shown in
%Figure~\ref{fig:Ldip}.

\begin{figure}
\begin{center}
\epsfig{file=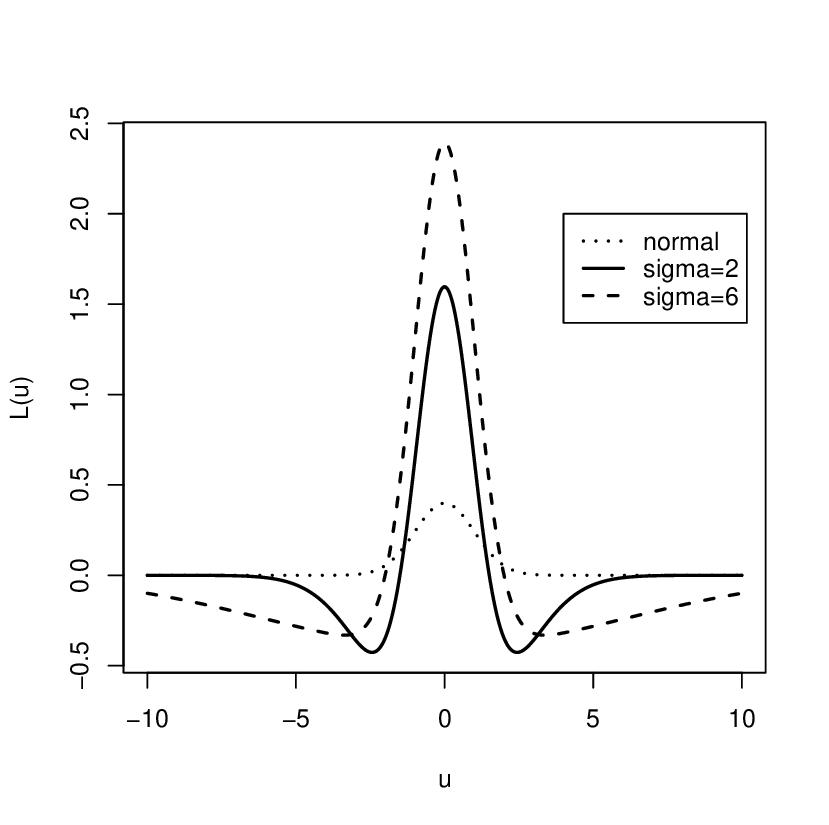,height=220pt} \caption{ Selection kernels in
$\mathcal L_3$. The dotted curve corresponds to the Gaussian kernel,
and each of the other kernels has $\alpha=6$. \label{fig:Lnegative}}
\end{center}
\end{figure}

The second family is $\mathcal
L_2=\bigl\{L(\cdot;\alpha,\sigma): \alpha>0,
\frac{\alpha}{1+\alpha}\leq\sigma\leq1\bigr\}$. Kernels in $\mathcal
L_2$ are densities which can be unimodal or bimodal. Note that the
Gaussian kernel is a member of this family.
The third sub-family is $\mathcal
L_3=\bigl\{L(\cdot;\alpha,\sigma): \alpha>0, \sigma>1\}$, each member
of which has negative tails. Examples of kernels in $\mathcal
L_3$ are shown in Figure~\ref{fig:Lnegative}.
%\begin{figure}
%\begin{center}
%\begin{tabular}{cc}
%{\bf(a)}&{\bf(b)}\\
%\epsfig{file=Lsigma0.5.eps,height=220pt}&\epsfig{file=Lalpha4.eps,height=220pt}\\
%\end{tabular}
%\caption{Selection kernels in $\mathcal L_1$. The dotted curve in both
%graphs corresponds to the Gaussian kernel. The other kernels in
%{\bf(a)} have $\sigma=0.5$ and the others in {\bf(b)} have
%$\alpha=4$. \label{fig:Ldip}}
%\end{center}
%\end{figure}

Kernels in  $\mathcal L_1$ and
$\mathcal L_3$ are not of the type usually used for estimating
$f$. Nonetheless, a worthwhile question is ``why not
use $L$ for both \cv {\it and} estimation of $f$?'' One could then
bypass the step of rescaling $\hatb$ and simply estimate $f$ by
an $L$-kernel estimator with bandwidth $\hatb$. The ironic answer to
this question is that the kernels in ${\cal L}$ that are best for \cv
purposes are very inefficient for estimating $f$. Indeed, it turns out
that an $L$-kernel estimator based on a sequence of ICV-optimal
kernels has $MISE$ that does not converge to 0 faster than $n^{-1/2}$. In
contrast, the $MISE$ of the best $\phi$-kernel estimator tends to 0
like $n^{-4/5}$. These facts fit with other \cv paradoxes, which
include the fact that LSCV outperforms other methods when the density
is highly structured, \citeN{Loader:Classical},
the improved performance of \cv in multivariate density
estimation, \citeN{Sainetal:cv}, and its improvement when the true density
is not smooth, \citeN{vanEs}. One could paraphrase these phenomena as
follows: ``The more difficult the function is to estimate, the better
\cv seems to perform.'' In our work, we have in essence made the
function more
difficult to estimate by using an inefficient kernel $L$. More details
on the $MISE$ of $L$-kernel estimators may be found
in~\citeN{Savchuk:thesis}.

%Given the intuition behind the TCV method
%discussed earlier and the ``cut-out-the middle'' nature of kernels in
%${\cal L}_1$, we initially surmised that ${\cal L}_1$ would be
%the best class for \cv purposes. However, it turns out that the best
%members of ${\cal
%  L}_3$ are at least marginally superior to those of ${\cal L}_1$ when
%used (as decribed in Section \ref{sec:meth}) to estimate $h_0$.

%\begin{equation}
%\label{eq:mixnorm}
%f(x)=\sum_{l=1}^k
%\frac{w_l}{\sigma_l}\phi\Bigl(\frac{x-\mu_l}{\sigma_l}\Bigr).
%$\end{equation}
%with $\sum_{l=1}^k w_l=1$.
%We rely on the class of the normal
% The corresponding expression
%for ISE~\eqref{eq:ISE} is the following:
%\[
%\begin{array}{rl}
%\ISE(h)=&\displaystyle{\frac{1}{n^2h\sqrt2}\sum_{i=1}^n\sum_{j=1}^n\phi\left(\frac{X_i-X_j}{h\sqrt2}\right)-
%\frac{2}{n}\sum_{i=1}^n\sum_{l=1}^k\frac{w_l}{\sqrt{\sigma_l^2+h^2}}\phi\left(\frac{X_i-\mu_l}{\sqrt{\sigma_l^2+h^2}}\right)}\\
%&+\displaystyle{\sum_{l_1=1}^k\sum_{l_2=1}^k\frac{w_{l_1}w_{l_2}}{\sqrt{\sigma_{l_1}^2+\sigma_{l_2}^2}}\phi\left(\frac{\mu_{l_1}-\mu_{l_2}}{\sqrt{\sigma_{l_1}^2+\sigma_{l_2}^2}}\right)}.
%\end{array}
%\]
\section{Large sample theory}\label{sec:theory}
%\subsection{ICV bandwidth and its statistical properties}
The theory presented in this section provides the underpinning for
our methodology. We first state a theorem on the asymptotic
distribution of $\hat h_{ICV}$, and then derive asymptotically optimal
choices for the parameters $\alpha$ and $\sigma$ of the selection kernel.

\subsection{Asymptotic mean squared error of the ICV
  bandwidth}\label{sec:MSE}
Classical theory of~\citeN{H&M:Extent} and~\citeN{Scott:UCV} entails that the
bias of an LSCV bandwidth is asymptotically
negligible in comparison to its standard deviation. We will show that
the variance of an ICV bandwidth can converge to 0 at a faster rate
than that of an LSCV bandwidth. This comes at the expense of a squared
bias that is {\it not} negligible. However, we will show how to select
$\alpha$ and $\sigma$ (the parameters of the selection kernel) so that the
variance and squared bias are balanced and the resulting mean squared
error tends to 0 at a faster rate than does that of the LSCV
bandwidth. The optimal rate of convergence of the relative error
$(\hat h_{ICV}-h_0)/h_0$ is $n^{-1/4}$, a substantial improvement over
the infamous $n^{-1/10}$ rate for LSCV.

Before stating our main result concerning the asymptotic distribution of
$\hat h_{ICV}$, we define some notation:
$$
\gamma(u)=\int L(w)L(w+u)\,du-2L(u),\quad \rho(u)=u\gamma'(u),
$$
$$
T_n(b)={\sum\sum}_{1\le i<j\le
  n}\left[\gamma\left(\frac{X_i-X_j}{b}\right)+\rho\left(\frac{X_i-X_j}{b}\right)\right],
$$
$$
T_n^{(j)}(b)=\frac{\partial^jT_n(b)}{\partial b^j},\quad j=1,2,
$$
$$
A_\alpha=\frac{3}{\sqrt{2\pi}}(1+\alpha)^2\left[\frac{1}{8}(1+\alpha)^2-\frac{8}{9\sqrt{3}}(1+\alpha)+\frac{1}{\sqrt2}\right],
$$
$$
C_\alpha=\frac{\sqrt{2A_\alpha}(2\sqrt{\pi})^{9/10}}{5(1+\alpha)^{9/5}\alpha^{1/5}}
\qquad {\rm and}\qquad D_\alpha=\frac{3}{20}\left(\frac{(1+\alpha)^2}{2\alpha^2\sqrt\pi}\right)^{2/5}.
$$
Note that to simplify notation, we have suppressed the fact that
$L$, $\gamma$ and $\rho$ depend on the parameters $\alpha$ and
$\sigma$. An outline of the proof of the following theorem is given in
the Appendix.

\noindent
{\bf Theorem.} \ {\sl Assume that $f$ and its first five derivatives are
continuous and bounded and that $f^{(6)}$ exists and is Lipschitz
continuous. Suppose also that
\begin{equation}\label{cond1}
(\hatb-b_0)\frac{T_n^{(2)}(\tilde b)}{T_n^{(1)}(b_0)}=o_p(1)
\end{equation}
for any sequence of random variables $\tilde b$ such that $|\tilde
b-b_0|\le|\hatb-b_0|$, a.s.
Then, if $\sigma=o(n)$ and $\alpha$ is fixed,
$$
\frac{\hat h_{ICV}-h_0}{h_0}=Z_nS_n+B_n+o_p(S_n+B_n),
$$
as $n\ra\infty$ and $\sigma\ra\infty$, where $Z_n$ converges in
distribution to a standard normal random variable,
\beq\label{eq:SD}
S_n=\left(\frac{1}{\sigma^{2/5}n^{1/10}}\right)
\frac{R(f)^{1/2}}{R(f'')^{1/10}}C_\alpha,
\eeq
and
\beq\label{eq:Bias}
B_n=\left(\frac{\sigma}{n}\right)^{2/5}\frac{R(f''')}{R(f'')^{7/5}}D_\alpha.
\eeq
}

%\bigi
\noindent
{\bf Remarks}

\medi
\begin{itemize}
\item[R1.] Assumption (\ref{cond1}) is only slightly stronger than assuming
  that $\hatb/b_0$ converges in probability to 1. To avoid making
  our paper overly technical we have chosen not to investigate sufficient
  conditions for (\ref{cond1}). However, this can be done using techniques as
  in~\citeN{Hall:CV} and~\citeN{H&M:Extent}.
\item[R2.] Theorem 4.1 of~\citeN{Scott:UCV} on asymptotic
  normality of LSCV bandwidths is not immediately
  applicable to our setting for at least three reasons: the kernel $L$
  is not   positive, it does not have compact support, and, most
  importantly,   it changes with $n$ via the parameter $\sigma$.
\item[R3.] The assumption of six derivatives for $f$ is required for a
  precise quantification of the asymptotic bias of $\hat
  h_{ICV}$. Our proof of asymptotic normality of $\hat b_{UCV}$ only
  requires that $f$ be four times differentiable, which coincides with
  the conditions of Theorem 4.1 in~\citeN{Scott:UCV}.
\item[R4.] The asymptotic bias $B_n$ is positive, implying that the ICV
  bandwidth tends to be larger than the optimal bandwidth. This is
  consistent with our experience in numerous simulations.
\end{itemize}
In the next section we apply the results of our theorem to determine
asymptotically optimal choices for $\alpha$ and $\sigma$.

\subsection{Minimizing asymptotic mean squared error}

The limiting distribution of $(\hat
h_{ICV}-h_0)/h_0$ has second moment $S_n^2+B_n^2$, where $S_n$ and
$B_n$ are defined by (\ref{eq:SD}) and (\ref{eq:Bias}).
%$$
%\left(\frac{1}{\sigma^{4/5}n^{1/5}}\right)C_\alpha^2\cdot\frac{R(f)}{R(f'')^{1/5}}+\left(\frac{\sigma}{n}\right)^{4/5}D_\alpha^2\cdot\frac{R(f''')^2}{R(f'')^{14/5}}.
%$$
Minimizing this expression with respect to $\sigma$ yields the
following asymptotically optimal choice for $\sigma$:
\beq\label{eq:sigopt}
\sigma_{n,opt}=n^{3/8}\left(\frac{C_\alpha}{D_\alpha}\right)^{5/4}\left[\frac{R(f)R(f'')^{13/5}}{R(f''')^2}\right]^{5/8}.
\eeq
The corresponding asymptotically optimal mean squared error is
\beq\label{eq:optmse}
MSE_{n,opt}=n^{-1/2}C_\alpha
D_\alpha\left[\frac{R(f''')R(f)^{1/2}}{R(f'')^{3/2}}\right],
\eeq
which confirms our previous claim that the relative error of $\hat
h_{ICV}$ converges to 0 at the rate $n^{-1/4}$. The corresponding
rates for LSCV and the Sheather-Jones plug-in rule are $n^{-1/10}$ and
$n^{-5/14}$, respectively.

Because $\alpha$ is not confounded with $f$ in $MSE_{n,opt}$, we may
determine a single optimal value of $\alpha$ that is independent of
$f$. The function $C_\alpha D_\alpha$ of $\alpha$ is
%$$
%C_\alpha D_\alpha=\frac{3\sqrt3}{50\cdot2^{1/4}}\left[\frac{1}{8}(1+\alpha)^2-\frac{8}{9\sqrt{3}}(1+\alpha)+\frac{1}{\sqrt2}\right]^{1/2}\alpha^{-1}.
%$$
%which is plotted in Figure~\ref{fig:conalpha}.
minimized at $\alpha_0=2.4233$. Furthermore, small
choices of $\alpha$ lead to an arbitrarily large increase in mean
squared error, while the MSE at $\alpha=\infty$ is only about 1.33
times that at the minimum.
%\begin{figure}
%\begin{center}
%\epsfig{file=conalpha.eps,height=230pt}
%\caption{How the asymptotically optimal MSE of $\hat h_{ICV}$ depends
%  on $\alpha$. The function plotted is $C_\alpha D_\alpha$, as defined
%  in expression (\ref{eq:optmse}). \label{fig:conalpha}}
%\end{center}
%\end{figure}

Our theory to this point applies to kernels in ${\cal L}_3$, i.e.,
kernels with negative tails. \citeN{Savchuk:thesis} has developed
similar theory for the case where $\sigma\ra0$,
which corresponds to $L\in {\cal L}_1$, i.e., kernels that apply negative
weights to the smallest spacings in the LSCV criterion. Interestingly,
the same optimal rate of $n^{-1/4}$ results from letting
$\sigma\ra0$. However, when the optimal values of $(\alpha,\sigma)$
are used in the respective cases ($\sigma\ra0$ and $\sigma\ra\infty$),
the limiting ratio of optimum mean squared errors is $0.752$, with
$\sigma\ra\infty$ yielding the smaller error. Our
simulation studies confirm that using $L$ with large $\sigma$ does
lead to more accurate estimation of the optimal bandwidth.

\section{Practical choice of $\alpha$ and $\sigma$\label{sec:MSEopt}}
In order to have an idea of how good choices of $\alpha$ and $\sigma$
vary with $n$ and $f$, we determined the minimizers of
the asymptotic mean squared error of $\hat h_{ICV}$ for various sample
sizes and densities. In doing so, we considered a single expression for
the asymptotic mean squared error that is valid for either large or
small values of $\sigma$. Furthermore, we
use a slightly enhanced version of
the asymptotic bias of $\hat h_{ICV}$. The first order bias of $\hat
h_{ICV}$ is $Cb_0-h_0$, or $C(b_0-b_n)+(h_n-h_0)$, where
\beq\label{eq:b_n}
b_n=\left(\frac{R(L)}{\mu_{2L}^2R(f'')}\right)^{1/5}n^{-1/5}\quad{\rm
  and}\quad
h_n=\left(\frac{R(\phi)}{\mu_{2\phi}^2R(f'')}\right)^{1/5}n^{-1/5}.
\eeq
Now, the term $h_n-h_0$ is of smaller order asymptotically than
$C(b_0-b_n)$ and hence was deleted in the theory of Section
\ref{sec:theory}. Here we retain $h_n-h_0$, and hence the
$\alpha$ that minimizes the mean squared error depends on both $n$ and
$f$.

We considered the following five normal
mixtures defined in the article by~\citeN{MarronWand:MISE}:

\begin{center}
\begin{tabular}{ll}
Gaussian density:&$N(0,1)$\\
Skewed unimodal density:&$\frac{1}{5}N(0,1)+\frac{1}{5}N\Bigl(\frac{1}{2},\bigl(\frac{2}{3}\bigr)^2\Bigr)+\frac{3}{5}N\Bigl(\frac{13}{12},\bigl(\frac{5}{9}\bigr)^2\Bigr)$\\
Bimodal
density:&$\frac{1}{2}N\Bigl(-1,\bigl(\frac{2}{3}\bigr)^2\Bigr)+\frac{1}{2}N\Bigl(1,\bigl(\frac{2}{3}\bigr)^2\Bigr)$\\
Separated bimodal
density:&$\frac{1}{2}N\Bigl(-\frac{3}{2},\bigl(\frac{1}{2}\bigr)^2\Bigr)+\frac{1}{2}N\Bigl(\frac{3}{2},\bigl(\frac{1}{2}\bigr)^2\Bigr)$\\
Skewed bimodal
density:&$\frac{3}{4}N(0,1)+\frac{1}{4}N\Bigl(\frac{3}{2},\bigl(\frac{1}{3}\bigr)^2\Bigr)$.\\
\end{tabular}
\end{center}
These choices for $f$ provide a fairly representative range of density
shapes. It is worth noting that the asymptotically optimal
$\sigma$ (expression (\ref{eq:sigopt})) is free of location and scale.
We may thus choose a single representative of a
location-scale family when investigating the effect of $f$.
The following remarks summarize our findings about $\alpha$ and $\sigma$.
\begin{itemize}
\item For each $n$, the optimal value of $\sigma$ ($\alpha$) is
  larger (smaller) for
  the unimodal densities than for the bimodal ones.
\item All of the MSE-optimal $\alpha$ and $\sigma$ correspond to
  kernels from $\mathcal L_3$, the family of negative-tailed kernels.
\item For each density, the optimal $\alpha$ decreases monotonically
  with $n$. Recall from Section 3.2 that the asymptotically optimal
  $\alpha$ is $2.42$. For each unimodal density, the optimal
  $\alpha$ is within 13.5\% of 2.42 at $n=1000$, and for each bimodal
  density is within 18\% of 2.42 when $n$ is 20,000.
\end{itemize}
%This does not support our
%original thought that the cut-out-the-middle kernels from $\mathcal
%L_1$ would be the best for use at the cross-validation stage.

In practice it would be desirable to have choices of $\alpha$ and
$\sigma$ that would adapt to the $n$ and $f$ at hand.
However, attempting to
estimate optimal values of $\alpha$ and $\sigma$ is potentially as
difficult as the bandwidth selection problem itself.
We have built a practical purpose model for $\alpha$ and $\sigma$ by
using polynomial regression. The independent variable was
$\log_{10}(n)$ and the dependent variables were the MSE-optimal
values of $\log_{10}(\alpha)$ and $\log_{10}(\sigma)$ for the five
densities defined above.
%The $\log_{10}$ transformations for $\alpha$
%and $\sigma$
%stabilize variability.
Using a sixth degree polynomial for
$\alpha$ and a quadratic for $\sigma$, we arrived at the following
models for $\alpha$ and $\sigma$:

\begin{equation}
\label{eq:model}
\begin{array}{l}
\alpha_{mod}=10^{3.390-1.093\log10(n)+0.025\log10(n)^3-0.00004\log10(n)^6}\\
\sigma_{mod}=10^{-0.58+0.386\log10(n)-0.012\log10(n)^2},\quad 100\leq n\leq500000.
\end{array}
\end{equation}

To the extent that unimodal densities are more prevalent than
multimodal densities in practice, these model values are biased
towards bimodal cases. Our extensive experience shows that the penalty
for using good bimodal choices for $\alpha$ and $\sigma$ when in fact
the density is unimodal, is an increase in the upward bias of $\hat
h_{ICV}$. Our
implementation of ICV, however, guards against oversmoothing by using
an objective upper bound on the bandwidth, as we explain in detail in
Section 7. We thus feel confident in recommending model (\ref{eq:model}) for
choosing $\alpha$ and $\sigma$ in practice, at least until a better
method is proposed. Indeed, this model is what we used to choose
$\alpha$ and $\sigma$ in the simulation study reported upon in Section
7.

\section{Robustness of ICV to data rounding\label{sec:Robust}}
\citeN[p.52]{Silverman:book} showed that if the data are
rounded to such an extent that the number of pairs $i<j$ for which
$X_i=X_j$ is above a threshold, then $LSCV(h)$
approaches $-\infty$ as $h$ approaches zero. This threshold is
$0.27n$ for the Gaussian kernel.
%Suppose the discretized data have
%$m$ ties.
\citeN{Chiu:discretization} showed that for data with ties, the
behavior of $LSCV(h)$ as $h\rightarrow 0$ is
determined by the balance between $R(K)$ and $2K(0)$. In particular,
$\lim_{h\ra0}LSCV(h)$ is $-\infty$ and $\infty$ when $R(K)<2K(0)$ and
$R(K)>2K(0)$, respectively. The former condition holds necessarily if $K$
is nonnegative and has its maximum at $0$. This means that all
the traditional kernels have the problem of choosing
$h=0$ when the data are rounded.

\begin{figure}
\begin{center}
\epsfig{file=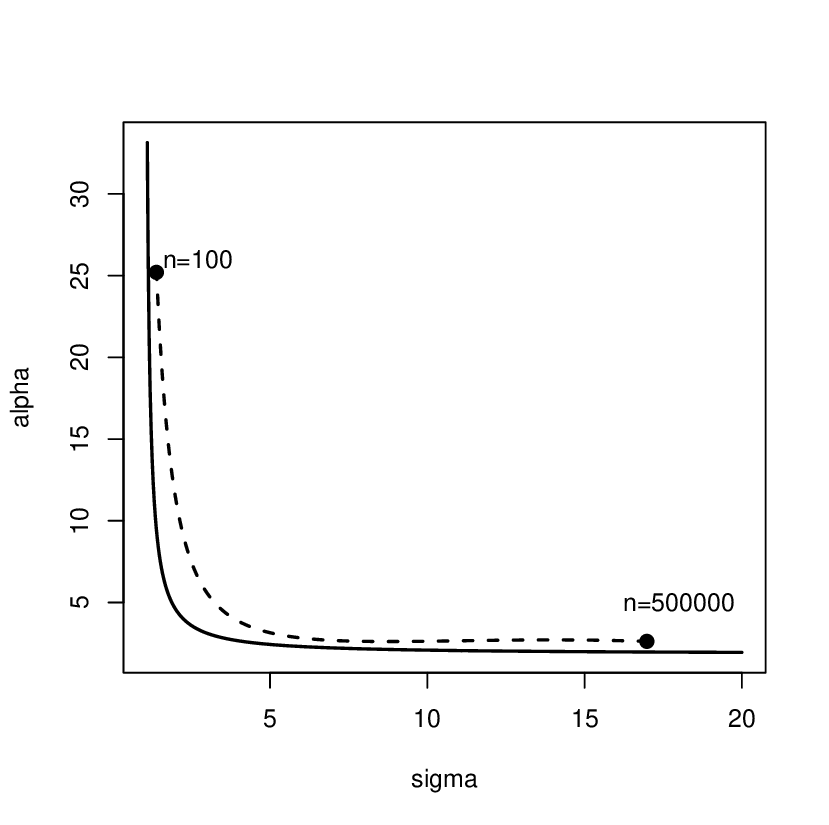,height=3in} \caption{Selection kernels
robust to rounding have $\alpha$ and $\sigma$ above the solid curve.
Dashed curve corresponds to the model-based selection
kernels.\label{fig:robustwild}}
\end{center}
\end{figure}

Recall that selection kernels~\eqref{eq:L} are not restricted to be
nonnegative. It turns out that there exist $\alpha$ and $\sigma$
such that $R(L)>2L(0)$ will hold. We say that selection kernels
satisfying this condition are robust to rounding. It can be
verified that the negative-tailed selection kernels with
$\sigma>1$ are robust to rounding when
\begin{equation}
\label{eq:roundalpha}
\alpha>\frac{-a_\sigma+\sqrt{a_\sigma+(2-1/\sqrt2)b_\sigma}}{b_\sigma},
\end{equation}
where
$a_\sigma=\left(\frac{1}{\sqrt2}-\frac{1}{\sqrt{1+\sigma^2}}-1+\frac{1}{\sigma}\right)$
and $b_\sigma=\left(\frac{1}{\sqrt2}-\frac{2}{\sqrt{1+\sigma^2}}+
\frac{1}{\sigma\sqrt2}\right)$.
It turns out that all the selection kernels corresponding to model
(\ref{eq:model}) are robust to rounding.
Figure~\ref{fig:robustwild} shows the
region~\eqref{eq:roundalpha} and also the curve defined by
model~\eqref{eq:model} for $100\leq n\leq 500000$.
Interestingly, the boundary
separating robust from nonrobust kernels almost coincides with the
$(\alpha,\sigma)$ pairs defined by that model.

\section{Local ICV}
A local version of cross-validation for density estimation was
proposed and analyzed independently by \citeN{HaSch} and
\citeN{MSV}. A local method allows the
bandwidth to vary with $x$, which is desirable when the
smoothness of the underlying density varies sufficiently with
$x$. \citeN{FHMP} proposed a different method of local smoothing that
is a hybrid of plug-in and cross-validation methods. Here we propose
that ICV be performed locally. The method parallels that of
\citeN{HaSch} and \citeN{MSV}, with the main difference being that
each local
bandwidth is chosen by ICV rather than LSCV.  We suggest using the
{\it smallest} local minimizer of the ICV curve, since ICV does not
have LSCV's tendency to undersmooth.

Let $\hat f_b$ be a kernel estimate that employs a kernel in
the class ${\cal L}$, and define, at the point $x$, a local ICV curve
by
$$
ICV(x,b)=\frac{1}{w}\int_{-\infty}^\infty\phi\left(\frac{x-u}{w}\right)\hat
f_b^2(u)\,du-\frac{2}{nw}\sum_{i=1}^n\phi\left(\frac{x-X_i}{w}\right)\hat
f_{b,-i}(X_i), \quad b>0.
$$
The quantity $w$ determines the degree to which the \cv
is local, with a very large choice of $w$ corresponding to global ICV.
Let $\hat b(x)$ be the minimizer of $ICV(x,b)$ with respect to
$b$. Then the bandwidth of a Gaussian kernel estimator at the point
$x$ is taken to be $\hat h(x)=C\hat b(x)$.  The constant $C$ is
defined by (\ref{eq:hnbn}), and choice of $\alpha$ and $\sigma$ in the
selection kernel will be discussed in Section 8.

Local LSCV can be criticized on the grounds that, at any $x$, it
promises to be even more unstable than global LSCV since it
(effectively) uses only a fraction of the $n$
observations. Because of its much greater stability, ICV seems to be a
much more feasible method of local bandwidth selection than does
LSCV. We provide evidence of this stability by example in Section 8.

\section{Simulation study \label{sec:Sims}}

The primary goal of our simulation study is to
compare ICV with ordinary LSCV. However, we will also include
the Sheather-Jones plug-in method in the study.
We considered the four sample sizes $n=100$, 250, 500 and 5000, and
sampled from each of the five densities listed in
Section~\ref{sec:MSEopt}. For each combination of density and
sample size, 1000 replications were performed. Here we give
only a synopsis of our results. The reader is referred to \citeN{SHS}
for a much more detailed account of what we observed.

Let $\hat h_0$ denote the minimizer of $ISE(h)$ for a Gaussian kernel
estimator. For each replication, we computed
%estimates of $E\bigl\{ISE(\hat h)/ISE(\hat h_0)\bigr\}$
%and ${\mbox{Median}}\bigl\{ISE(\hat h)/ISE(\hat h_0)\bigr\}$
$\hat h_0$, $\hat h_{ICV}^*$, $\hat h_{UCV}$ and $\hat
h_{SJPI}$. The definition of $\hat h_{ICV}^*$ is $\min(\hat
h_{ICV},\hat h_{OS})$, where $\hat h_{OS}$  is the oversmoothed
bandwidth of \citeN{overTerr}. Since $\hat h_{ICV}$ tends to be biased
upwards, this is a convenient means of limiting the bias.
In all cases the parameters  $\alpha$ and $\sigma$ in the
selection kernel $L$ were chosen according to model (\ref{eq:model}).
For any random variable $Y$ defined in each replication of our
simulation, we denote the average of $Y$ over all replications
(with $n$ and $f$ fixed) by $\widehat E(Y)$. Our main conclusions may
be summarized as follows.
\begin{itemize}
\item The ratio $\widehat E(\hat h_{ICV}^*-\widehat E\hat h_0)^2/\widehat
  E(\hat h_{UCV}-\widehat E\hat h_0)^2$ ranged between 0.04 and 0.70
  in the sixteen settings excluding the skewed bimodal
  density. For the skewed bimodal, the ratio was 0.84, 1.27, 1.09,
  and 0.40 at the
  respective sample sizes 100, 250, 500 and 5000. The fact that this
  ratio was larger than 1 in two cases was a result of ICV's bias, since
  the sample standard deviation of the ICV bandwidth was smaller than
  that for the LSCV bandwidth in all twenty settings.
\item The ratio $\widehat{E}\bigl(ISE(\hat{h}_{ICV}^*)/ISE(\hat h_0)\bigr)/
\widehat{E}\bigl(ISE(\hat{h}_{UCV})/ISE(\hat h_0)\bigr)$ was smaller
than 1 for every combination of density and sample
size. For the two ``large bias'' cases mentioned in the previous
remark the ratio  was 0.92.
\item The ratio $\widehat{E}\bigl(ISE(\hat{h}_{ICV}^*)/ISE(\hat h_0)\bigr)/
\widehat{E}\bigl(ISE(\hat{h}_{SJPI})/ISE(\hat h_0)\bigr)$ was smaller
than 1 in six of the twenty cases considered. Among the other fourteen
cases, the ratio was between 1.00 and 1.15, exceeding 1.07 just twice.
\item Despite the fact that the LSCV bandwidth is asymptotically normally
distributed (see~\citeN{H&M:Extent}), its distribution in finite
samples tends to be skewed to the left. In contrast, our simulations
show that the ICV bandwidth distribution is nearly symmetric.
\end{itemize}

%$\widehat{\mbox{Median}}(\hat h)$ the average and the median of $\hat h$,
%respectively, computed over all replications.
%We also computed the performance measure $\widehat
%E(\hat h-\hat h_0)^2$, the average squared distance of $\hat h$ from
%$\hat h_0$.

\section{Examples}

In this Section we illustrate the use of ICV with two examples, one
involving credit scores from Fannie Mae and the other simulated data.
The first example is provided to compare the
ICV, LSCV, and Sheather-Jones plug-in methods for choosing a global
bandwidth. The second example illustrates the benefit of applying
ICV locally.

\subsection{Mortgage defaulters}

In this example we analyze the credit scores of Fannie Mae clients
who defaulted on their loans. The mortgages considered were purchased
in ``bulk''
lots by Fannie Mae from primary banking institutions. The data set
was taken from the website \url{http://www.dataminingbook.com}
associated with~\citeN{Shmueli:book}.

\begin{figure}[h]
\begin{center}
\begin{tabular}{cc}
\epsfig{file=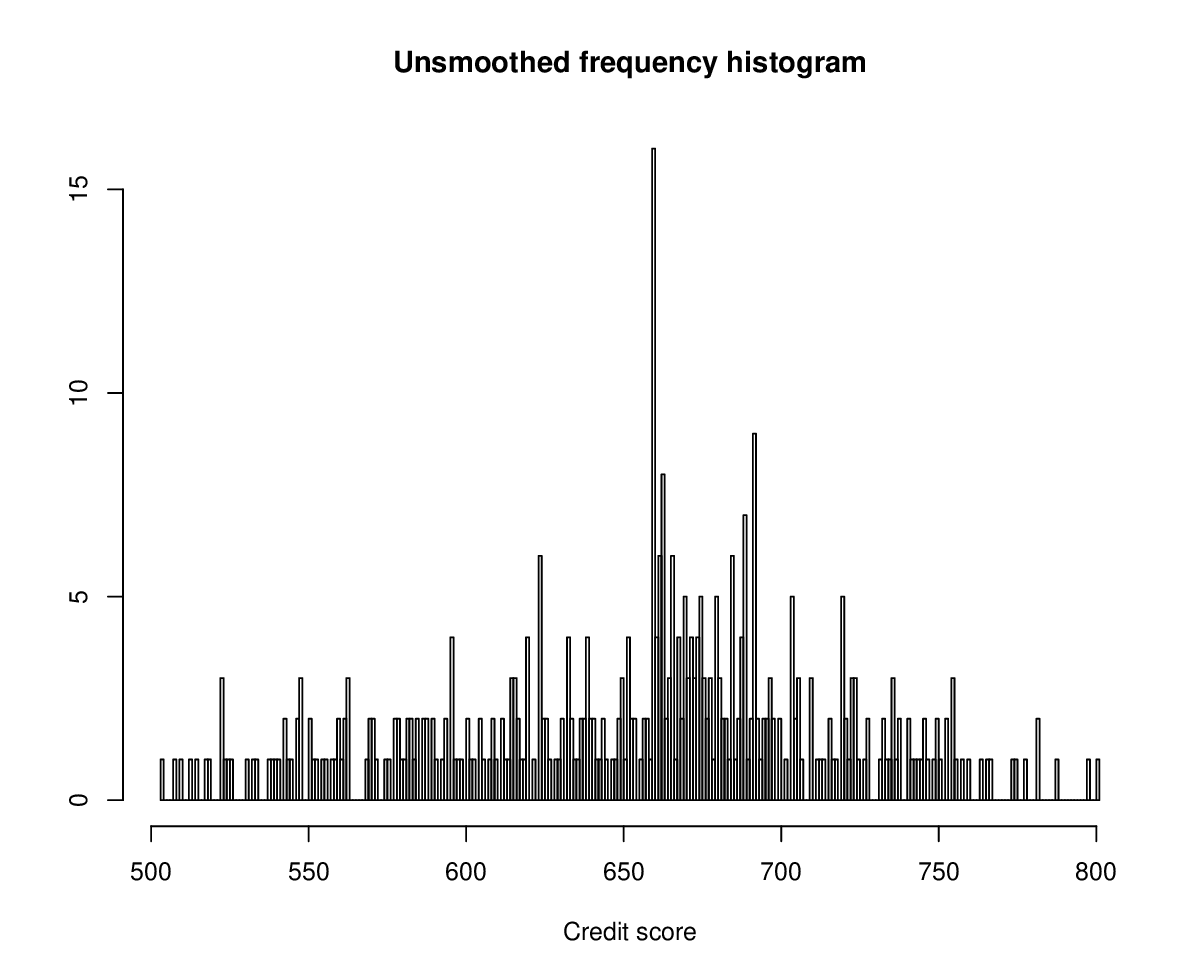,height=165pt}&\epsfig{file=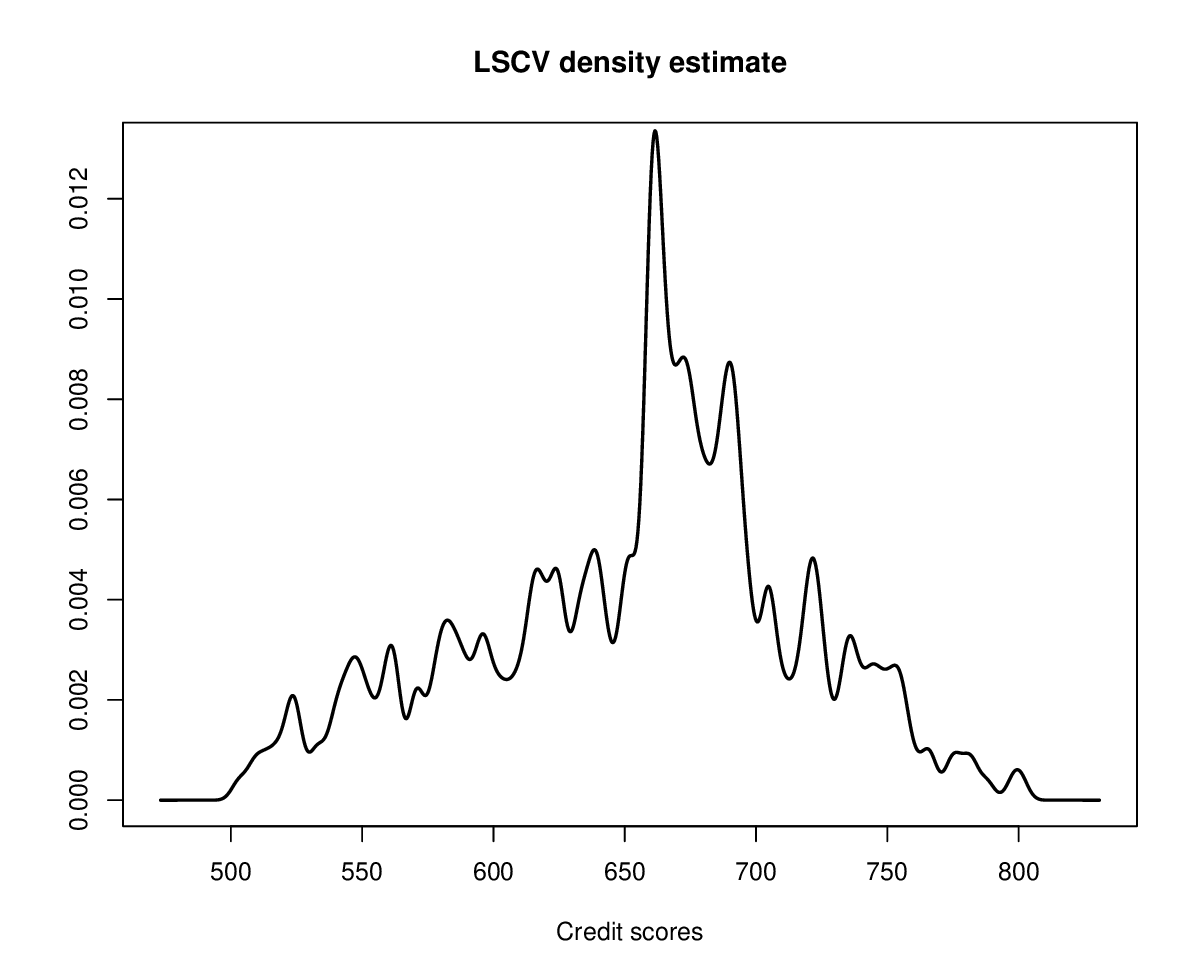,height=165pt}\\
\epsfig{file=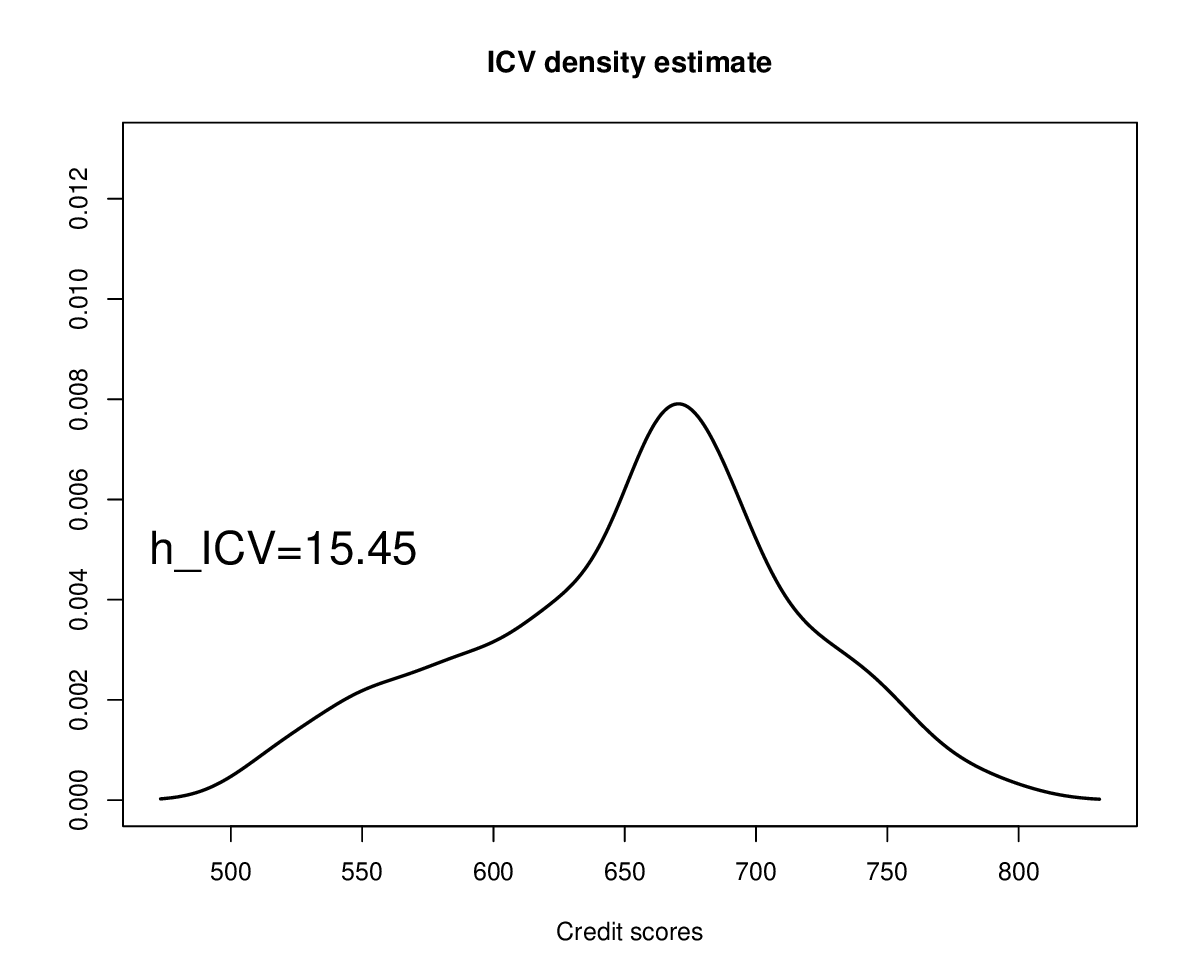,height=165pt}&\epsfig{file=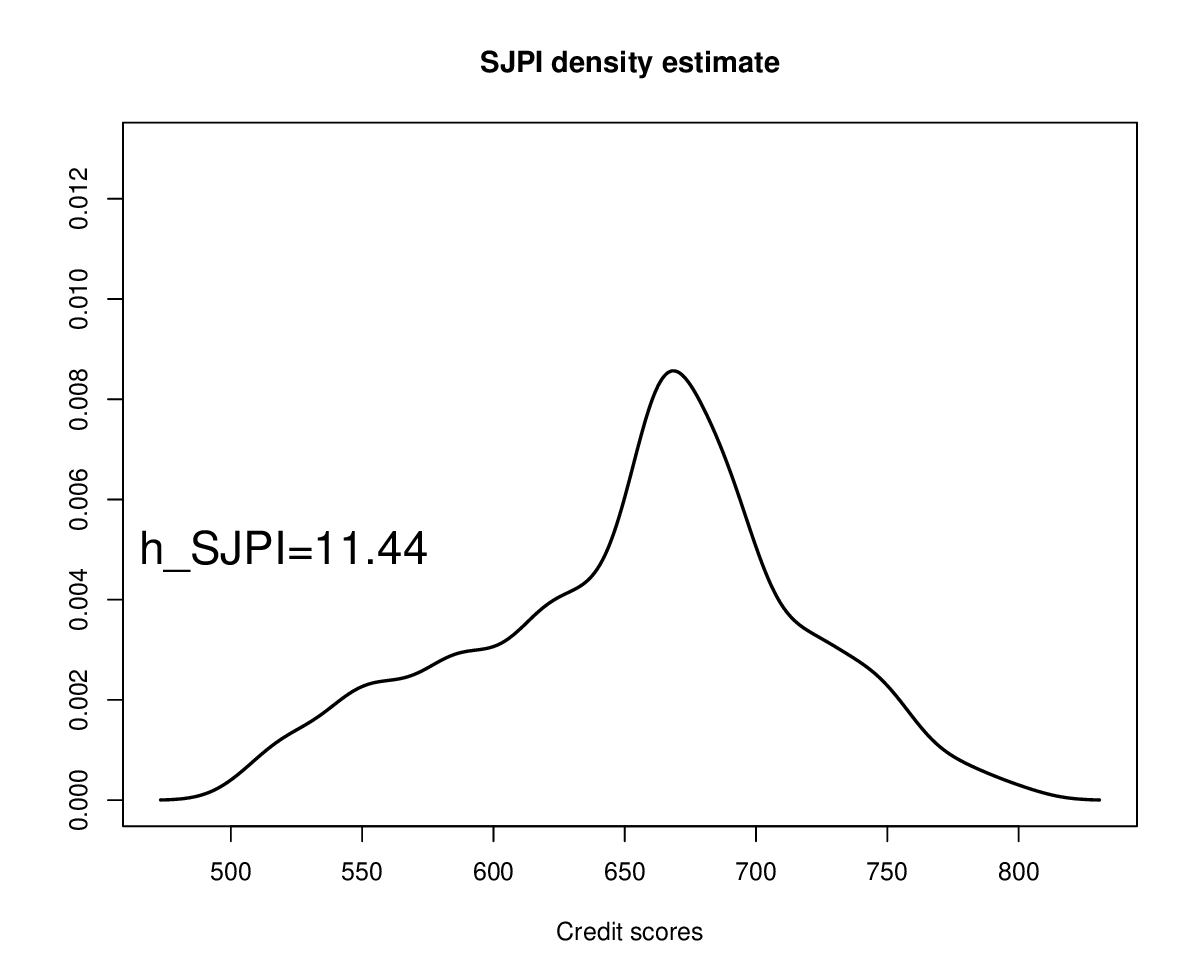,height=165pt}\\
\end{tabular}
\caption{Unsmoothed histogram and kernel density estimates for
credit scores. \label{fig:KDE_Mortgage}}
\end{center}
\end{figure}

In Figure~\ref{fig:KDE_Mortgage} we have plotted an unsmoothed
frequency histogram and the LSCV, ICV and Sheather-Jones plug-in
density estimates for the credit scores. The class interval size in
the unsmoothed histogram was chosen to be 1, which is equal to the
accuracy to which the data have been reported. It turns out that
the LSCV curve tends to $-\infty$ when $h\rightarrow 0$, but has
a local minimum at about 2.84. Using $h=2.84$ results in a severely
undersmoothed estimate. Both the Sheather-Jones plug-in and ICV density
estimates show a single mode around 675 and look similar, with the
ICV estimate being somewhat smoother.
Interestingly, a high percentage of the defaulters have
credit scores less than 620, which many lenders consider the minimum
score that qualifies for a loan; see \citeN{Desmond}.

\subsection{Local ICV: simulated example}

For this example we took five samples of size $n=1500$ from the
kurtotic unimodal density defined in~\citeN{MarronWand:MISE}. First,
we note that even the bandwidth that minimizes $ISE(h)$
results in a density estimate that is
much too wiggly in the tails. On the other hand, using local versions
of either ICV or LSCV resulted in much better density estimates, with
local ICV producing in each case a visually better estimate than
that produced by local LSCV.

For the local LSCV and ICV methods we considered four values of
$w$ ranging from 0.05 to 0.3. A selection kernel
with $\alpha=6$ and $\sigma=6$ was used in local ICV. This
$(\alpha,\sigma)$ choice performs well for global bandwidth selection
when the density is unimodal, and hence seems reasonable for local
bandwidth selection since locally the density should have relatively
few features. For a
given $w$, the local ICV and LSCV bandwidths were found for
$x=-3,-2.9,\ldots,2.9,3$, and were interpolated at other $x\in[-3,3]$
using a spline. Average
squared error (ASE) was used to measure closeness of a local
density estimate $\hat f_\ell$ to the true density $f$:
\[
ASE=\frac{1}{61}\sum_{i=1}^{61}(\hat f_\ell(x_i)-f(x_i))^2.
\]

\begin{figure}[h]
\begin{center}
\begin{tabular}{|cc|}
\hline
\multicolumn{2}{|c|}{{\large $\eta=0.05$}}\\
\epsfig{file=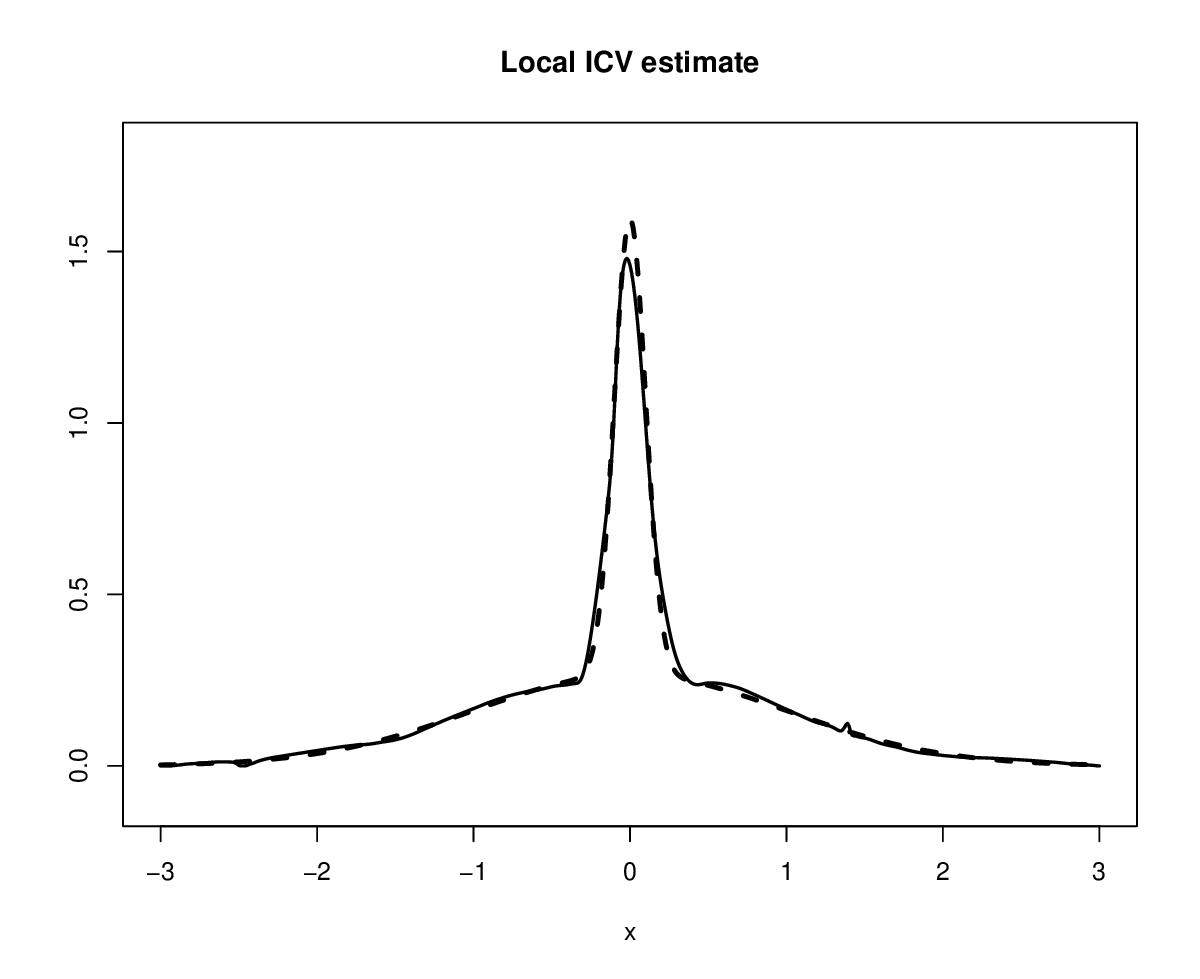,height=165pt}&\epsfig{file=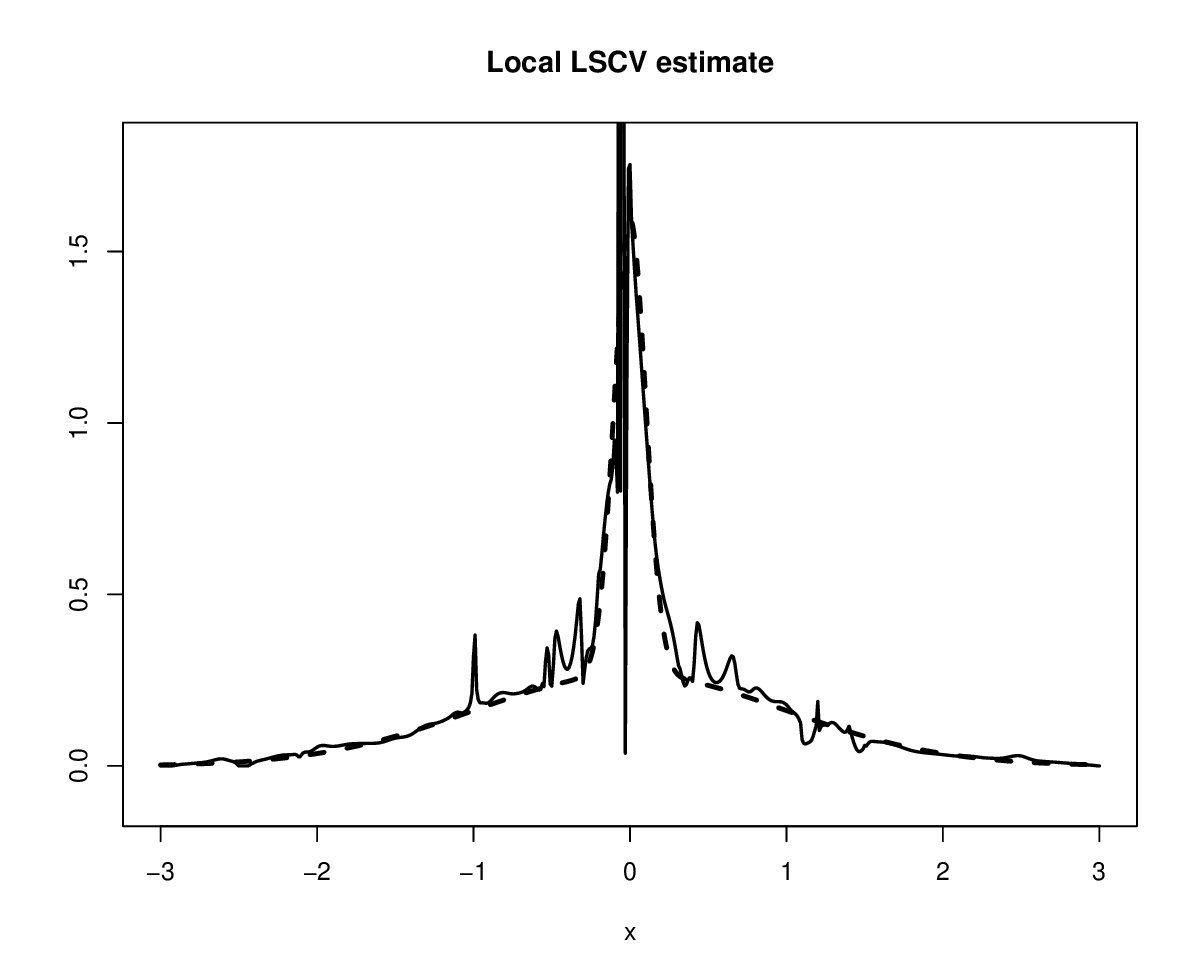,height=165pt}\\
$ASE=0.000778$&$ASE=0.001859$\\
\hline \hline
\multicolumn{2}{|c|}{{\large $\eta=0.3$}}\\
\epsfig{file=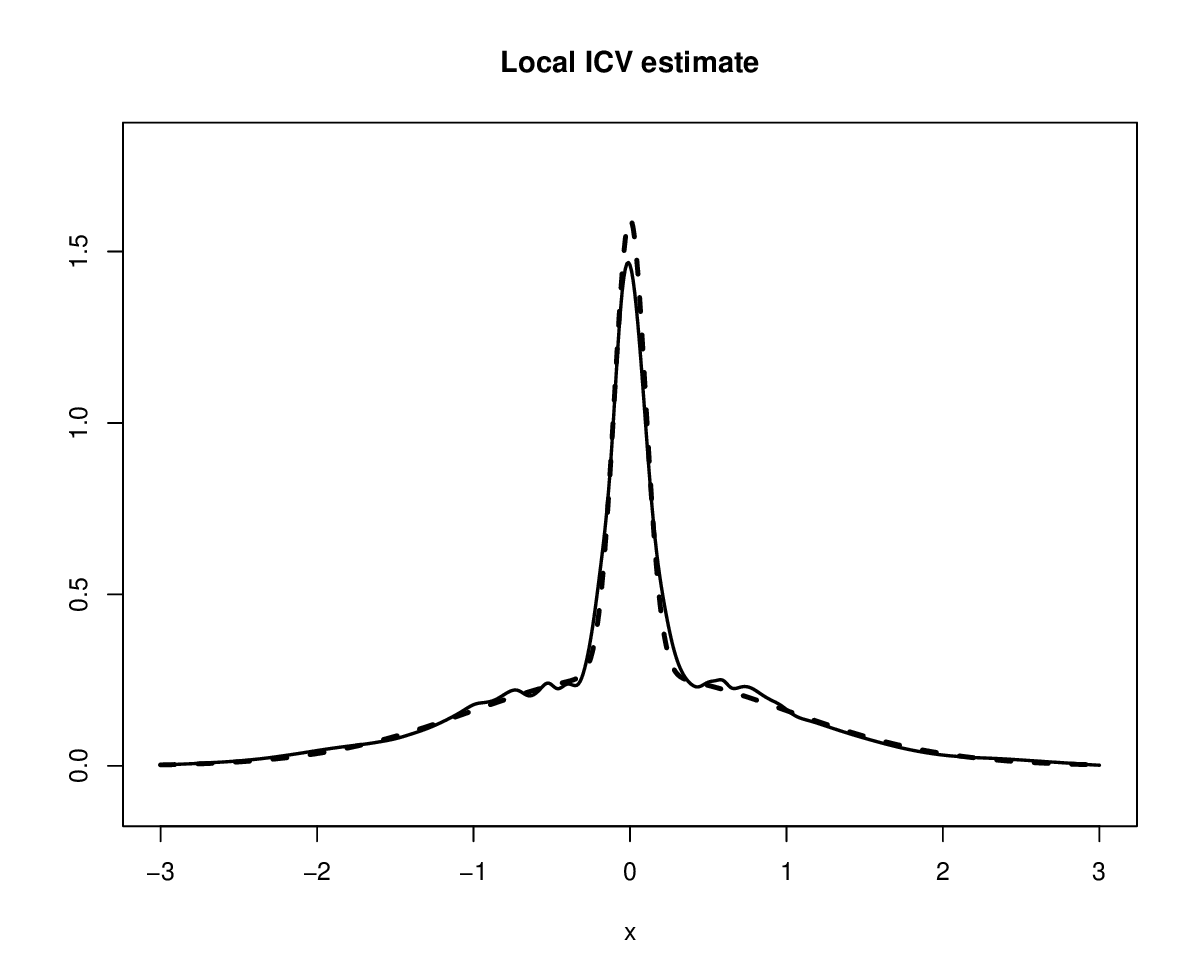,height=165pt}&\epsfig{file=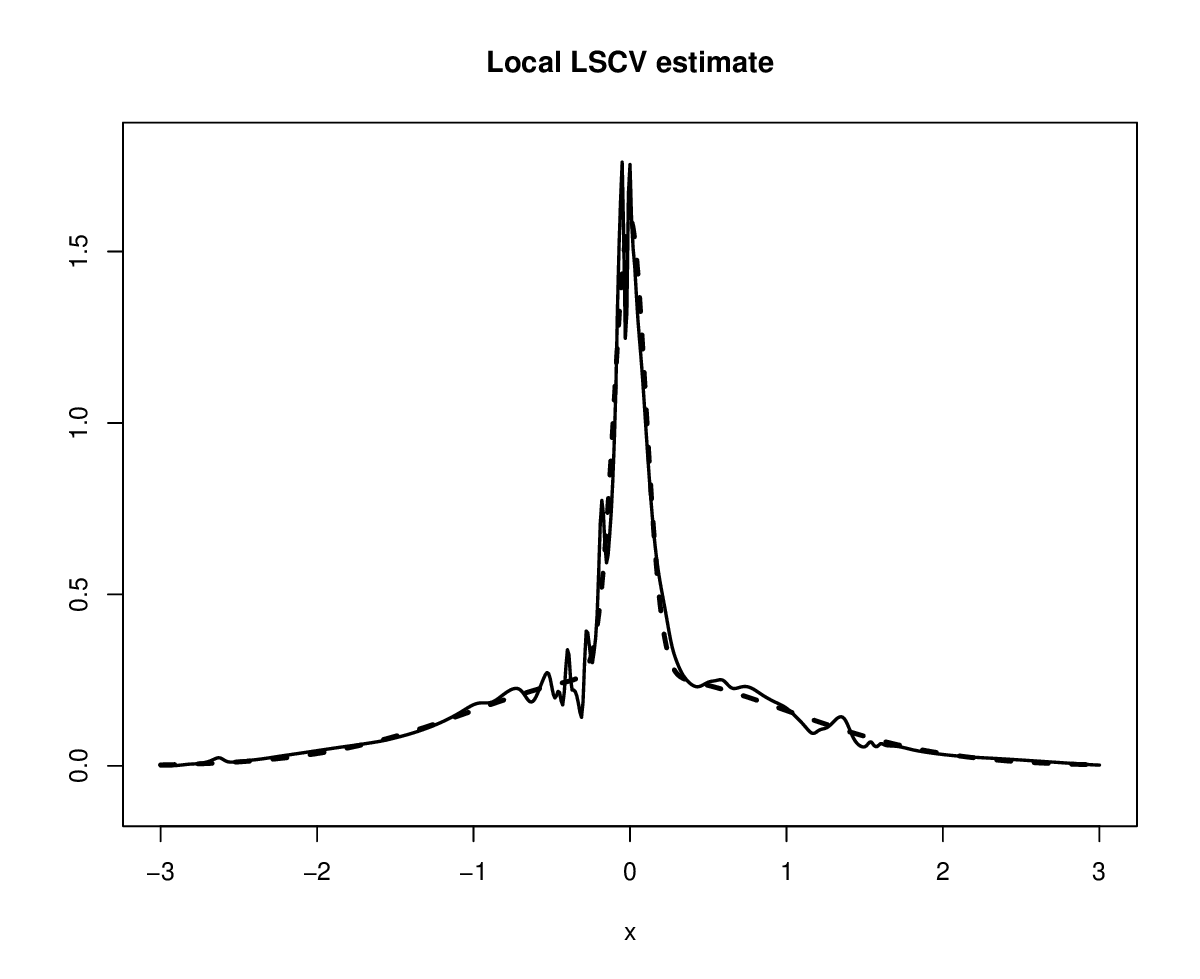,height=165pt}\\
$ASE=0.000762$&$ASE=0.001481$\\
\hline
\end{tabular}
\caption{The solid curves correspond to the local LSCV and ICV
density estimates, whereas the dashed curves show the kurtotic
unimodal density. \label{fig:Loc_estimates}}
\end{center}
\end{figure}

Figure~\ref{fig:Loc_estimates} shows results for one of the five
samples. Estimates corresponding to the smallest and the largest
values of $w$ are provided. The local ICV
method performed similarly well for all values of
$w$ considered, whereas all the local LSCV estimates were very
unsmooth, albeit with some improvement in smoothness as $w$ increased.

%\clearpage
\section{Summary}
A widely held view is that kernel choice is not terribly important
when it comes to estimation of the underlying curve. In this paper we
have shown that kernel choice can
have a dramatic effect on the properties of \cvp. Cross-validating
kernel estimates that use Gaussian or other traditional kernels
results in highly variable bandwidths, a result that has been
well-known since at least 1987. We have shown that certain
kernels with low efficiency for estimating $f$ can produce
cross-validation bandwidths whose relative error converges to 0 at a
faster rate than that of Gaussian-kernel cross-validation
bandwidths.

The kernels we have studied have the form
$(1+\alpha)\phi(u)-\alpha\phi(u/\sigma)/\sigma$, where
$\phi$ is the standard normal density and $\alpha$ and $\sigma$ are
positive constants. The interesting selection kernels in this class are
of two types: unimodal, negative-tailed kernels and ``cut-out the middle
kernels,'' i.e., bimodal kernels that go negative between the modes.
Both types of kernels yield the rate
improvement mentioned in the previous paragraph. However, the best
negative-tailed kernels yield bandwidths with smaller asymptotic mean
squared error than do the best ``cut-out-the-middle'' kernels.

A model
for choosing the selection kernel parameters has been
developed. Use of this model makes our method completely automatic.
%The
%parameters are ones that produce an asymptotically optimal rate of
%convergence for the ICV bandwidth.
A simulation study and
examples reveal that use of this method leads to improved performance
relative to ordinary LSCV.
%We have also derived a bias correction for
%ICV that performed well in our simulations.

To date we have considered only selection
kernels that are a linear combination of two normal densities. It is
entirely possible that another class of kernels would work even
better. In particular, a question of at least theoretical interest is
whether or not the convergence rate of $n^{-1/4}$ for the relative
bandwidth error can be improved upon.

\section{Appendix}
Here we outline the proof of our theorem in Section
\ref{sec:theory}. A much more detailed proof is available from the
authors.

We start by writing
\beqn
T_n(b_0)&=&T_n(\hatb)+(b_0-\hatb)T_n^{(1)}(b_0)+\frac{1}{2}
(b_0-\hatb)^2T_n^{(2)}(\tilde b)\\
        &=&-nR(L)/2+(b_0-\hatb)T_n^{(1)}(b_0)+\frac{1}{2}
        (b_0-\hatb)^2T_n^{(2)}(\tilde
b),
\eeqn
where $\tilde b$ is between $b_0$ and $\hatb$, and so
$$
(\hatb-b_0)\lp1-(\hatb-b_0)\frac{T_n^{(2)}(\tilde
  b)}{2T_n^{(1)}(b_0)}\rp=\frac{T_n(b_0)+nR(L)/2}{-T_n^{(1)}(b_0)}.
$$
Using condition (\ref{cond1}) we may write the last equation as
\beq\label{eq:delta}
(\hatb-b_0)=\frac{T_n(b_0)+nR(L)/2}{-T_n^{(1)}(b_0)}+o_p\lp\frac{T_n(b_0)+nR(L)/2}{-T_n^{(1)}(b_0)}\rp.
\eeq

Defining $s_n^2=\Var(T_n(b_0))$ and $\beta_n=E(T_n(b_0))+nR(L)/2$, we have
$$
\frac{T_n(b_0)+nR(L)/2}{-T_n^{(1)}(
  b_0)}=\frac{T_n(b_0)-ET_n(b_0)}{s_n}\cdot\frac{s_n}{-T_n^{(1)}(b_0)}+\frac{\beta_n}{-T_n^{(1)}(b_0)}.
$$
Using the central limit theorem of~\citeN{Hall:CLT}, it can be verified that
$$
Z_n\equiv\frac{T_n(b_0)-ET_n(b_0)}{s_n}\cond N(0,1).
$$
Computation of the first two moments of $T_n^{(1)}(b_0)$ reveals that
$$
\frac{-T_n^{(1)}(b_0)}{5R(f'')b_0^4\mu_{2L}^2n^2/2}\conp 1,
$$
and so
$$
\frac{T_n(b_0)+nR(L)/2}{-T_n^{(1)}(
  b_0)}=Z_n\cdot
\frac{2s_n}{5R(f'')b_0^4\mu_{2L}^2n^2}
+\frac{2\beta_n}{5R(f'')b_0^4\mu_{2L}^2n^2}+
o_p\lp \frac{s_n+\beta_n}{b_0^4\mu_{2L}^2n^2}\rp.
$$

At this point we need the first two moments of $T_n(b_0)$. A fact that
will be used frequently from this point on is that
$\mu_{2k,L}=O(\sigma^{2k})$, $k=1,2,\ldots$. Using our assumptions on the
smoothness of $f$, Taylor series expansions,
symmetry of $\gamma$ about 0 and $\mu_{2\gamma}=0$,
$$
ET_n(b_0)=-\frac{n^2}{12}b_0^5\mu_{4\gamma}R(f'')+\frac{n^2}{240}b_0^7\mu_{6\gamma}R(f''')+O(n^2b_0^8\sigma^7).
$$
Recalling the definition of $b_n$ from (\ref{eq:b_n}), we have
\beq\label{eq:ET}
\beta_n&=&-\frac{n^2}{12}b_0^5\mu_{4\gamma}R(f'')+\frac{n^2}{240}b_0^7\mu_{6\gamma}R(f''')\notag\\
&&+\frac{n^2}{2}b_n^5\mu_{2L}^2R(f'')+O(n^2b_0^8\sigma^7).
\eeq
Let $MISE_L(b)$ denote the MISE of an $L$-kernel estimator with
bandwidth $b$. Then
$MISE_L'(b_n)=(b_n-b_0)MISE_L''(b_0)+o\left[(b_n-b_0)MISE_L''(b_0)\right]$, implying that
\beq\label{eq:secord}
b_n^5=b_0^5+5b_0^4\frac{MISE_L'(b_n)}{MISE_L''(b_0)}+o\left[b_0^4\frac{MISE_L'(b_n)}{MISE_L''(b_0)}\right].
\eeq
Using a second order approximation to $MISE'_L(b)$ and a first order
approximation to $MISE''_L(b)$, we then have
$$
b_n^5=b_0^5-b_0^7\frac{\mu_{2L}\mu_{4L}R(f''')}{4\mu_{2L}^2R(f'')}+o(b_0^7\sigma^2).
$$
Substitution of this expression for $b_n$ into (\ref{eq:ET}) and using
the facts $\mu_{4\gamma}=6\mu_{2L}^2$,
$\mu_{6\gamma}=30\mu_{2L}\mu_{4L}$ and
$b_0\sigma=o(1)$, it follows that
$\beta_n=o(n^2b_0^7\sigma^6)$. Later in the proof we will
see that this last result implies that the first order bias of
$\hat h_{ICV}$ is due only to the difference $Cb_0-h_0$.

Tedious but straightforward calculations show that $s_n^2\sim
n^2b_0R(f)A_\alpha/2$, where $A_\alpha$ is as defined in Section
\ref{sec:MSE}.
It is worth noting that $A_\alpha=R(\rho_\alpha)$, where
$\rho_\alpha(u)=u\gamma'_\alpha(u)$ and
$\gamma_\alpha(u)=(1+\alpha)^2\int\phi(u+v)\phi(v)\,dv-2(1+\alpha)\phi(u)$.
One would expect from Theorem 4.1 of~\citeN{Scott:UCV} that
the factor $R(\rho)$ would appear in $\Var(T_n(b_0))$. Indeed it does
implicitly, since $R(\rho_\alpha)\sim R(\rho)$ as
$\sigma\ra\infty$. Our point is that, when $\sigma\ra\infty$,
the part of $L$ depending on $\sigma$ is negligible in terms of its
effect on $R(\rho)$ and also $R(L)$.

To complete the proof write
\beqn
\frac{\hat h_{ICV}-h_0}{h_0}&=&\frac{\hat
  h_{ICV}-h_0}{h_n}+o_p\left[\frac{\hat h_{ICV}-h_0}{h_n}\right]\\
&=&\frac{\hat
  b_{UCV}-b_0}{b_n}+\frac{(Cb_0-h_0)}{h_n}+o_p\left[\frac{\hat
    h_{ICV}-h_0}{h_n}\right].
\eeqn
Applying the same approximation of $b_0$ that led to
(\ref{eq:secord}), and the analogous one for $h_0$, we have
\beqn
\frac{Cb_0-h_0}{h_n}&=&b_n^2\frac{\mu_{2L}\mu_{4L}R(f''')}{20\mu_{2L}^2R(f'')}-h_n^2\frac{\mu_{2\phi}\mu_{4\phi}R(f''')}{20\mu_{2\phi}^2R(f'')}+o(b_n^2\sigma^2+h_n^2)\\
&=&\frac{R(L)^{2/5}\mu_{2L}\mu_{4L}R(f''')}{20(\mu_{2L}^2)^{7/5}R(f'')^{7/5}}\,n^{-2/5}+o(b_n^2\sigma^2).
\eeqn
It is easily verified that, as $\sigma\ra\infty$, $R(L)\sim
(1+\alpha)^2/(2\sqrt\pi)$, $\mu_{2L}\sim-\alpha\sigma^2$ and
$\mu_{4L}\sim -3\alpha\sigma^4$, and hence
$$
\frac{Cb_0-h_0}{h_n}=\lp\frac{\sigma}{n}\rp^{2/5}\frac{R(f''')}{R(f'')^{7/5}}D_\alpha+o\left[\lp\frac{\sigma}{n}\rp^{2/5}\right].
$$
The proof is now complete upon combining all the previous results.
\section{Acknowledgements}
The authors are grateful to  David Scott and George Terrell for
providing valuable insight about \cvp, and to three referees and an
associate editor, whose comments led to a much improved final version
of our paper. The research of Savchuk and Hart was supported in part
by NSF Grant DMS-0604801.

\bibliographystyle{chicago}

\begin{thebibliography}{}

\bibitem[\protect\citeauthoryear{Ahmad and Ran}{Ahmad and Ran}{2004}]{Ahmad}
Ahmad, I.~A. and I.~S. Ran (2004).
\newblock Kernel contrasts: a data-based method of choosing smoothing
  parameters in nonparametric density estimation.
\newblock {\em J. Nonparametr. Stat.\/}~{\em 16\/}(5), 671--707.

\bibitem[\protect\citeauthoryear{Bowman}{Bowman}{1984}]{Bowman:LSCV}
Bowman, A.~W. (1984).
\newblock An alternative method of cross-validation for the smoothing of
  density estimates.
\newblock {\em Biometrika\/}~{\em 71\/}(2), 353--360.

\bibitem[\protect\citeauthoryear{Chiu}{Chiu}{1991a}]{Chiu:CV}
Chiu, S.-T. (1991a).
\newblock Bandwidth selection for kernel density estimation.
\newblock {\em Ann. Statist.\/}~{\em 19\/}(4), 1883--1905.

\bibitem[\protect\citeauthoryear{Chiu}{Chiu}{1991b}]{Chiu:discretization}
Chiu, S.-T. (1991b).
\newblock The effect of discretization error on bandwith selection for kernel
  density estimation.
\newblock {\em Biometrika\/}~{\em 78\/}(2), 436--441.

\bibitem[\protect\citeauthoryear{Desmond}{Desmond}{2008}]{Desmond}
Desmond, M. (2008).
\newblock Lipstick on a pig.
\newblock {\em Forbes\/}.

\bibitem[\protect\citeauthoryear{Fan, Hall, Martin, and Patil}{Fan
  et~al.}{1996}]{FHMP}
Fan, J., P.~Hall, M.~A. Martin, and P.~Patil (1996).
\newblock On local smoothing of nonparametric curve estimators.
\newblock {\em J. Amer. Statist. Assoc.\/}~{\em 91\/}(433), 258--266.

\bibitem[\protect\citeauthoryear{Feluch and Koronacki}{Feluch and
  Koronacki}{1992}]{Feluch:spacing}
Feluch, W. and J.~Koronacki (1992).
\newblock A note on modified cross-validation in density estimation.
\newblock {\em Comput. Statist. Data Anal.\/}~{\em 13\/}(2), 143--151.

\bibitem[\protect\citeauthoryear{Hall}{Hall}{1983}]{Hall:CV}
Hall, P. (1983).
\newblock Large sample optimality of least squares cross-validation in density
  estimation.
\newblock {\em Ann. Statist.\/}~{\em 11\/}(4), 1156--1174.

\bibitem[\protect\citeauthoryear{Hall}{Hall}{1984}]{Hall:CLT}
Hall, P. (1984).
\newblock Central limit theorem for integrated square error of multivariate
  nonparametric density estimators.
\newblock {\em J. Multivariate Anal.\/}~{\em 14\/}(1), 1--16.

\bibitem[\protect\citeauthoryear{Hall and Marron}{Hall and
  Marron}{1987}]{H&M:Extent}
Hall, P. and J.~S. Marron (1987).
\newblock Extent to which least-squares cross-validation minimises integrated
  square error in nonparametric density estimation.
\newblock {\em Probab. Theory Related Fields\/}~{\em 74\/}(4), 567--581.

\bibitem[\protect\citeauthoryear{Hall and Schucany}{Hall and
  Schucany}{1989}]{HaSch}
Hall, P. and W.~R. Schucany (1989).
\newblock A local cross-validation algorithm.
\newblock {\em Statist. Probab. Lett.\/}~{\em 8\/}(2), 109--117.

\bibitem[\protect\citeauthoryear{Hart and Yi}{Hart and Yi}{1998}]{HartYi}
Hart, J.~D. and S.~Yi (1998).
\newblock One-sided cross-validation.
\newblock {\em J. Amer. Statist. Assoc.\/}~{\em 93\/}(442), 620--631.

\bibitem[\protect\citeauthoryear{Jones, Marron, and Sheather}{Jones
  et~al.}{1996}]{Jones:survey}
Jones, M.~C., J.~S. Marron, and S.~J. Sheather (1996).
\newblock A brief survey of bandwidth selection for density estimation.
\newblock {\em J. Amer. Statist. Assoc.\/}~{\em 91\/}(433), 401--407.

\bibitem[\protect\citeauthoryear{Loader}{Loader}{1999}]{Loader:Classical}
Loader, C.~R. (1999).
\newblock Bandwidth selection: classical or plug-in?
\newblock {\em Ann. Statist.\/}~{\em 27\/}(2), 415--438.

\bibitem[\protect\citeauthoryear{Marron and Wand}{Marron and
  Wand}{1992}]{MarronWand:MISE}
Marron, J.~S. and M.~P. Wand (1992).
\newblock Exact mean integrated squared error.
\newblock {\em Ann. Statist.\/}~{\em 20\/}(2), 712--736.

\bibitem[\protect\citeauthoryear{Mielniczuk, Sarda, and Vieu}{Mielniczuk
  et~al.}{1989}]{MSV}
Mielniczuk, J., P.~Sarda, and P.~Vieu (1989).
\newblock Local data-driven bandwidth choice for density estimation.
\newblock {\em J. Statist. Plann. Inference\/}~{\em 23\/}(1), 53--69.

\bibitem[\protect\citeauthoryear{Rudemo}{Rudemo}{1982}]{Rudemo:LSCV}
Rudemo, M. (1982).
\newblock Empirical choice of histograms and kernel density estimators.
\newblock {\em Scand. J. Statist.\/}~{\em 9\/}(2), 65--78.

\bibitem[\protect\citeauthoryear{Sain, Baggerly, and Scott}{Sain
  et~al.}{1994}]{Sainetal:cv}
Sain, S.~R., K.~A. Baggerly, and D.~W. Scott (1994).
\newblock Cross-validation of multivariate densities.
\newblock {\em J. Amer. Statist. Assoc.\/}~{\em 89\/}(427), 807--817.

\bibitem[\protect\citeauthoryear{Savchuk}{Savchuk}{2009}]{Savchuk:thesis}
Savchuk, O. (2009).
\newblock {\em Choosing a kernel for cross-validation}.
\newblock Ph{D} thesis, Texas A\&M University.

\bibitem[\protect\citeauthoryear{Savchuk, Hart, and Sheather}{Savchuk
  et~al.}{2008}]{SHS}
Savchuk, O.~Y., J.~D. Hart, and S.~J. Sheather (2008).
\newblock An empirical study of indirect cross-validation.
\newblock {\em Festschrift for Tom Hettmansperger. IMS Lecture Notes-Monograph
  Series\/}.
\newblock Submitted.

\bibitem[\protect\citeauthoryear{Scott and Terrell}{Scott and
  Terrell}{1987}]{Scott:UCV}
Scott, D.~W. and G.~R. Terrell (1987).
\newblock Biased and unbiased cross-validation in density estimation.
\newblock {\em J. Amer. Statist. Assoc.\/}~{\em 82\/}(400), 1131--1146.

\bibitem[\protect\citeauthoryear{Sheather and Jones}{Sheather and
  Jones}{1991}]{Sheather:PI}
Sheather, S.~J. and M.~C. Jones (1991).
\newblock A reliable data-based bandwidth selection method for kernel density
  estimation.
\newblock {\em J. Roy. Statist. Soc. Ser. B\/}~{\em 53\/}(3), 683--690.

\bibitem[\protect\citeauthoryear{Shmueli, Patel, and Bruce}{Shmueli
  et~al.}{2006}]{Shmueli:book}
Shmueli, G., N.~R. Patel, and P.~C. Bruce (2006).
\newblock {\em Data Mining for Business Intelligence: Concepts, Techniques, and
  Applications in Microsoft Office Excel with XLMiner}.
\newblock New York: Wiley.

\bibitem[\protect\citeauthoryear{Silverman}{Silverman}{1986}]{Silverman:book}
Silverman, B.~W. (1986).
\newblock {\em Density estimation for statistics and data analysis}.
\newblock Monographs on Statistics and Applied Probability. London: Chapman \&
  Hall.

\bibitem[\protect\citeauthoryear{Stute}{Stute}{1992}]{Stute}
Stute, W. (1992).
\newblock Modified cross-validation in density estimation.
\newblock {\em J. Statist. Plann. Inference\/}~{\em 30\/}(3), 293--305.

\bibitem[\protect\citeauthoryear{Terrell}{Terrell}{1990}]{overTerr}
Terrell, G.~R. (1990).
\newblock The maximal smoothing principle in density estimation.
\newblock {\em J. Amer. Statist. Assoc.\/}~{\em 85\/}(410), 470--477.

\bibitem[\protect\citeauthoryear{van Es}{van Es}{1992}]{vanEs}
van Es, B. (1992).
\newblock Asymptotics for least squares cross-validation bandwidths in
  nonsmooth cases.
\newblock {\em Ann. Statist.\/}~{\em 20\/}(3), 1647--1657.

\end{thebibliography}

\end{document}